\documentclass[twocolumn,aps,showpacs,pra]{revtex4}
\usepackage{epsfig}
\usepackage{psfrag}

\newcommand{\ket}[1]{| #1 \rangle}
\newcommand{\bra}[1]{\langle #1 |}

\newcommand{\Caption}[1]{\caption{ #1}}

\newcommand{\br}{{\bf r}}

\begin{document}

\title{The GPE and higher order theories in one-dimensional Bose gases}
\author{M.D.~Lee} 
\affiliation{Clarendon Laboratory, Department of Physics, University of Oxford,
Parks Road, Oxford OX1 3PU, United Kingdom} 
\author{S.A.~Morgan} 
\affiliation{Department of Physics and Astronomy, University College London, 
Gower Street, London WC1E~6BT, U.K.} 
\author{K.~Burnett} 
\affiliation{Clarendon Laboratory, Department of Physics, University of Oxford,
Parks Road, Oxford OX1 3PU, United Kingdom}

\date{\today}

\begin{abstract} 
We investigate the properties of the one-dimensional Bose gas at zero
temperature, for which exact results exist for some model systems.  We treat
the interactions between particles in the gas with an approximate form of the
many-body T-matrix, and find a form of Gross-Pitaevskii equation valid in 1D. 
The results presented agree with the exact models in both the weakly and
strongly interacting limits, and interpolate smoothly between them. We also
investigate the use of mean-field treatments of trapped BECs to describe the 1D
system in the strongly interacting limit. We find that the use of the many-body
T-matrix to describe interactions leads to qualitative agreement with the exact
models for some physical quantities.  We indicate how the standard mean-field
treatments need to be modified to extend and improve the agreement.
\end{abstract}

\maketitle

\section{Introduction}

Recent advances in experimental techniques~\cite{Gorlitz2001a,Schreck2001a}
have stimulated interest in the topic of ultra-cold dilute Bose gases in one
dimension.  The one-dimensional (1D) case leads to some interesting physics
which does not occur in higher dimensions, and is also of theoretical
importance due to the existence of exactly soluble
models~\cite{Girardeau1960a,Lieb1963a,Lieb1963b}.  In this paper we investigate
the 1D Bose gas in the limit of zero temperature.  We extend our earlier
work~\cite{Lee2002a} on the interactions in two dimensions to the 1D case, and
obtain a 1D form of Gross-Pitaevskii equation (GPE).  We also apply standard
methods of mean-field theory developed to describe three-dimensional
Bose-Einstein condensates (BECs) to the 1D trapped case.  In both cases we
compare the results obtained from these methods to the exact solutions.

It is well known that a true BEC cannot occur at any finite temperature in an
interacting homogeneous Bose gas in 1D~\cite{Hohenberg1967a}, due to long
wavelength fluctuations; nor can it occur in the limit of zero
temperature~\cite{Pitaevskii1991a}, due to quantum fluctuations. However, the
presence of a trapping potential changes the density of states at low energies,
and in the weakly-interacting limit a BEC may be formed~\cite{Ketterle1996a}. 
In a recent paper, Petrov \emph{et al.\ }discussed three different regimes of
quantum degeneracy which can occur in a  confined 1D interacting
system~\cite{Petrov2000b}: BEC, quasi-condensate, and Tonks-Girardeau gas.  In
the weakly interacting limit a BEC exists, but as the interactions become
stronger the mean-field energy becomes important.  When this energy becomes of
the order of the energy level spacing of the trap, fluctuations again become
significant.  In this regime, the system forms a quasi-condensate, with local
phase coherence, rather than the global coherence associated with a true BEC.
In this regime the gas is similar to the homogeneous case, in which it was
shown that at $T=0$ the phase coherence has only a power-law decay, rather than
the exponential decay occurring at higher temperatures~\cite{Schwartz1977a}.
Finally, in the limit of very strong interactions, the particles become
impenetrable and the system is known as a Tonks-Girardeau
gas~\cite{Tonks1936a,Girardeau1960a}.  This limit is exactly soluble, due to a
mapping relationship between the wave function for the bosons and that of a
non-interacting fermion gas~\cite{Girardeau1960a}.

In the following section we briefly outline the relevant exact results for a 1D
gas.  Section~\ref{sec:1DGPE} then presents a form of the GPE in which we use
an approximate many-body T-matrix to describe the interactions in a 1D gas. 
The GPE gives the correct results in both the weakly-interacting and
Tonks-Girardeau limits, and interpolates smoothly between these regimes.  We
confirm that the non-linear dependence in the GPE changes from cubic in the
weakly-interacting regime to quintic in the strongly-interacting regime, and
that this arises due to the many-body effects on the scattering of particles.
In Sec.~\ref{sec:highorder1D}, we then discuss several standard mean-field
theories and present the results of these for the trapped 1D gas. In the weakly
interacting limit these theories should be valid since a true BEC exists in
this regime.  Our aim is to push several different theories into the
strongly-interacting limit with the dual objectives of describing the 1D system
and investigating the strengths and weaknesses of these theories.

\section{Exact results in 1D}

The 1D homogeneous system of $N$ bosons which interact via a delta-function
potential of the form
\begin{equation}
V(x_1-x_2) = V_0\delta(x_1-x_2),
\end{equation}
is exactly integrable.  In the limit $V_0 \rightarrow \infty$ the particles are
impenetrable (the Tonks-Girardeau gas), and exact results can be obtained even
in spatially inhomogeneous systems.  The exact solutions are found by applying
the appropriate boundary conditions to the wave function describing the system.

In the impenetrable limit the wave function must necessarily vanish wherever
two particles meet (i.e. wherever $x_i=x_j$).  Girardeau showed
in~\cite{Girardeau1960a} that the exact $N$-boson ground state wave function
subject to such a boundary condition is equal to $|\psi_F(x_1,...,x_N)|$, where
$\psi_F(x_1,...,x_N)$ is the wave function of $N$ non-interacting spinless
fermions governed by the single-particle Hamiltonian $\hat{H}_{\rm sp}$
(including terms from any external potential).  Since the wave functions of
such fermions must be antisymmetric with respect to co-ordinate exchanges,
their wave functions automatically vanish whenever $x_i=x_j$, and so the
Tonks-Girardeau gas boundary condition is implicitly obeyed by such a gas.  

A number of bulk properties of a 1D Tonks-Girardeau gas may be calculated quite
trivially as a result of this Fermi-Bose mapping
theorem~\cite{Girardeau1960a,Girardeau2001a}.  For example, the ground state
energy of a Tonks-Girardeau gas is simply found from a summation of the
energies of the first $N$ non-interacting single-particle states (i.e. those
that would be occupied by $N$ ideal fermions in the ground state
configuration). Thus, for a homogeneous gas~\cite{Girardeau1960a}
\begin{equation}
{E_0 \over N} = {\hbar^2 \over m} {\pi^2 \over 6} n^2,
\label{eq:tonkshomogE}
\end{equation}
where $n$ is the density.  The chemical potential is then easily
calculated as
\begin{equation}
\mu = {\partial E_0 \over \partial N} = {\hbar^2 \over m}{\pi^2 \over 2} n^2.
\label{eq:tonkshomogmu}
\end{equation}
Similarly, the chemical potential for a Tonks-Girardeau gas in a harmonic
potential of frequency $\omega$ is
\begin{equation}
\mu = \hbar\omega(N+1/2).
\label{eq:tonksmu}
\end{equation}

The system of $N$ bosons interacting with an arbitrary value of $V_0$ was
solved in 1963 by Lieb and Liniger~\cite{Lieb1963a,Lieb1963b} for a homogeneous
gas.  The Tonks-Girardeau gas results are recovered in the limit $V_0
\rightarrow \infty$.  The ground-state energy of the Lieb-Liniger gas is given
(at $T=0$) by
\begin{equation}
{E_0 \over N} = {\hbar^2 \over 2m} n^2 e(\gamma), \label{eq:LLenergy}
\end{equation}
where $e(\gamma)$ is a solution to the Lieb-Liniger system of equations, and is
a function of the dimensionless quantity $\gamma = V_0m/\hbar^2n$ which was
found to be the important parameter in the system.  The function $e(\gamma)$
has the following asymptotic behaviour
\begin{equation}
e(\gamma) \approx \left\{ \begin{array}{ll}
\gamma & \mbox{  for }\gamma \rightarrow 0,\\
{\pi^2\over 3}\left( {\gamma \over \gamma+2 }\right)  
& \mbox{  for }\gamma \rightarrow \infty,
\end{array}\right.
\end{equation}
and the energy per particle agrees with Eq.~(\ref{eq:tonkshomogE}) in the
impenetrable limit.  The chemical potential is given by the equation
\begin{equation}
\mu = {\hbar^2 \over 2m} n^2\left[3e(\gamma) - \gamma {de(\gamma)\over d\gamma}
\right].
\end{equation}

In the $\gamma \rightarrow 0$ weakly-interacting limit, the exact results were
found to be in agreement with the predictions of standard Bogoliubov theory. 
The Bogoliubov predictions begin to fail as $\gamma$ is increased and for
$\gamma \gtrsim 2$ the results do not agree well with the exact
theory~\cite{Lieb1963a}.  Interestingly, for fixed $V_0$ the $\gamma
\rightarrow 0$ limit corresponds to the high-density limit, whilst a high value
of $\gamma$ implies a small density.  This is then the reverse of the 3D case
where Bogoliubov theory is valid in the low density limit.  In a subsequent
section we will show to what extent higher-order modifications to Bogoliubov
theory can improve the agreement for higher $\gamma$.

For a system where we keep the density constant, and increase the interaction
strength $V_0$ from the non-interacting limit to the impenetrable limit (i.e.
increasing $\gamma$), we expect the chemical potential to vary in the following
manner.  In the low-$V_0$ limit the chemical potential increases linearly with
$V_0$, but as $V_0$ is increased further $\mu$ starts to saturate, and in the
high-$V_0$ limit it becomes independent of $V_0$, in agreement with
Eq.~(\ref{eq:tonkshomogmu}).  In the following section we introduce a modified
GPE which follows this behaviour.

\section{The GPE in 1D}
\label{sec:1DGPE}

We consider the model system of a gas of one-dimensional point particles
interacting with an arbitrary strength $V_0$, corresponding to the Lieb-Liniger
gas.  We emphasize here that we are interested in the limit in which the
interparticle interactions in the system are genuinely one-dimensional, because
it is only in this limit that we can compare to the exact results mentioned
earlier.

We can write a one-dimensional Gross-Pitaevskii equation for a system
of $N$ particles by a straightforward generalisation
of the usual 3D equation, giving
\begin{equation}
\mu\psi(x) = \left[-{\hbar^2 \over 2m}{d^2 \over dx^2} + 
V_{\rm trap}(x) + g_{\rm
1D}N|\psi(x)|^2\right]\psi(x),
\label{eq:1DGPE}
\end{equation}
where $\psi(x)$ is the condensate wave function, and $V_{\rm trap}(x)$
represents the trapping potential.  The coupling parameter $g_{\rm 1D}$
describes the interactions between particles, and we discuss the nature of this
quantity in the following section before presenting the results predicted by
the GPE.

\subsection{Interactions and the GPE}

The GPE can easily be derived by functional differentiation of the Hamiltonian
for a gas with a general interparticle interaction $V(x)$.  The coupling
parameter is then given by the matrix element $\bra{0}V(x)\ket{0}$, where the
particles in both the incoming and outgoing states have zero relative momentum
(see, for example,~\cite{Morgan2000a}).  However, a representation in terms of
matrix elements of the bare interparticle potential is undesirable, since for a
realistic potential they are often very large and frequently the exact
potential itself is unknown.  In order to avoid this it is common for 3D
problems to use the two-body T-matrix $T_{\rm 2B}$ in place of the
interparticle potential.  This T-matrix describes the scattering of two
particles in a vacuum, and it is possible to rewrite the interaction term in
the Hamiltonian in terms of $T_{\rm 2B}$~\cite{Morgan2000a}.  Carrying out the
functional differentiation again leads to a GPE of the above form, where the
coupling parameter is now $g = \bra{0}T_{\rm 2B}(E=0)\ket{0}$.  Here $E$ is the
energy of the interacting particles, and for two condensate particles this is
zero.  In three dimensions this leads to the familiar result $g_{\rm 3D} =
4\pi\hbar^2a/m$, where $a$ is the \emph{s}-wave scattering length.  However, in
1D (and 2D) the two-body T-matrix vanishes in the zero energy and momentum
limit, and has significant imaginary terms at a finite real energy (see, for
example,~\cite{Morgan2002a}).

In earlier papers~\cite{Lee2002a,Lee2002b}, we have argued that the many-body
effects of the surrounding atoms on particle interactions, while being a small
perturbation in 3D, are crucial in lower dimensions.  The many-body effects can
be incorporated into a many-body T-matrix $T_{\rm
MB}$~\cite{Stoof1996a,Bijlsma1997a,Proukakis1998a,Morgan2000a}, and the
coupling parameter in the GPE becomes
\begin{equation}
g_{\rm 1D} = \bra{0}T_{\rm MB}(E=0)\ket{0}.
\end{equation}
Unfortunately, the many-body T-matrix is
very difficult to solve exactly.  However, in~\cite{Lee2002a} we presented an
approximate solution, valid in the zero temperature limit, in which the
many-body T-matrix was approximated by the two-body T-matrix evaluated at a
shifted effective interaction energy (off the energy shell).  The GPE in 1D is
therefore given by Eq.~(\ref{eq:1DGPE}) with the coupling parameter
\begin{equation}
g_{\rm 1D} = \bra{0}T_{\rm MB}(E=0)\ket{0} 
\approx \bra{0}T_{\rm 2B}(E^*)\ket{0},
\label{eq:gapprox}
\end{equation}
where $E^*$ is the shifted interaction energy, and was found in~\cite{Lee2002a}
to be negative.  As an illustration of the utility and accuracy of this
approximation, we discuss its application to the 3D case in the appendix.

\subsection{The hard-core limit}

We will look first at the GPE in the impenetrable, $V_0 \rightarrow \infty$,
limit in which the system is a Tonks-Girardeau gas.  The off-shell two-body
T-matrix for such a 1D gas, calculated in~\cite{Morgan2002a}, is
\begin{eqnarray}
\lefteqn{\bra{0} T_{\rm 2B}^{\rm Tonks}(E) \ket{0} = } \nonumber \\ 
&&\left\{ \begin{array}{ll}
-\frac{2\hbar^2}{ma} \left( i\sqrt{mE \over \hbar^2}a + 
{mE \over \hbar^2}a^2 \right) &\mbox{for } E>0, \\
-\frac{2\hbar^2}{ma} \left( \sqrt{m|E| \over \hbar^2}a - 
{mE \over \hbar^2}a^2 \right) &\mbox{for }E<0,
\end{array}\right. ,
\label{eq:offshellT2Btonks}
\end{eqnarray}
where $a$ is the size of the impenetrable particle, and for the delta-function
gas we take the limit that $a \rightarrow 0$. Combined with
Eq.~(\ref{eq:gapprox}) this gives a coupling parameter of the
form~\cite{Lee2002a}
\begin{equation}
g_{\rm Tonks}(\mu) =  \sqrt{4C\hbar^2 \mu/m}
\label{eq:gtonkssqrtmu}
\end{equation}

to leading order for a homogeneous system. Here we have taken $E^* =-C\mu$ and
assumed that $C>0$~\cite{Lee2002a}, (note that if this is not the case then the
coupling parameter is imaginary to leading order).  Using the homogeneous
solution to the GPE $\mu = g_{\rm Tonks}n$ we can rewrite the coupling
parameter in terms of the density $n$ as
\begin{equation}
g_{\rm Tonks}(n) = {4C\hbar^2 \over m}n.
\end{equation}
Combining this with $\mu = g_{\rm Tonks}n$ gives $\mu \propto n^2$, and a
comparison with the exact result found by Girardeau and given in
Eq.~(\ref{eq:tonkshomogmu}), shows that we have obtained the correct functional
dependence of $\mu$ on $n$.  This stems from the dependence of the coupling
parameter on $\sqrt{\mu}$ in Eq.~(\ref{eq:gtonkssqrtmu}).  This is further
verification of our approximation for the many-body T-matrix in terms of the
off-shell two-body T-matrix.  A comparison with the Girardeau result also
shows that $C = \pi^2/8$.  

To apply this result to a trapped system, we use a local density approximation,
assuming that the scattering between two particles occurs on a length scale
much shorter than the variation in the potential.  First we solve the
homogeneous expression $\mu = g_{\rm Tonks}(\mu)n$ to obtain an expression
for the chemical potential in terms of the density $n$.  Applying the local
density approximation then requires $\mu(n(x))  = g_{\rm Tonks}(x)n(x)$ to
hold for all $x$, and we obtain a spatially-dependent coupling parameter
\begin{equation}
g_{\rm Tonks}(x) =  {\pi^2\hbar^2 N \over 2 m} |\psi(x)|^2,
\label{eq:1Dspatialgtonks}
\end{equation}
where we have used $n(x) = N|\psi(x)|^2$.

We can now write the GPE for a 1D trapped condensate as
\begin{equation}
\mu\psi(x) =\left[-{\hbar^2 \over 2m}{d^2 \over dx^2} + 
V_{\rm trap}(x)+ {\pi^2\hbar^2 N^2 \over 2m}|\psi(x)|^4\right]\psi(x).
\label{eq:tonksGPE}
\end{equation}
This GPE differs significantly from the 3D version by virtue of the fact that
the nonlinear term is now proportional to $|\psi|^4\psi$ instead of
$|\psi|^2\psi$.  In the Thomas-Fermi limit we can neglect the kinetic energy
term and the solution for the wave function can easily be shown to be
\begin{equation}
|\psi(x)| = {\left(\mu - V_{\rm trap}(x)\right)^{1/4} 
\over (\hbar^2\pi^2N^2/2m)^{1/4}}
\theta \left(\mu - V_{\rm trap}(x)\right)
\label{eq:quinticGPETF}
\end{equation}
This differs substantially from the 3D and 2D cases, where the Thomas-Fermi
wave function is instead proportional to $(\mu - V_{\rm trap}(x))^{1/2}$, due
to the different form of the non-linearity.

The Tonks-Girardeau GPE  and the results presented in this section are in exact
agreement with the earlier work of Kolomeisky \emph{et al.\
}\cite{Kolomeisky2000a}, who obtained the same form of coupling parameter from
a renormalisation group approach~\cite{Kolomeisky1992a}\footnote{Note that in
the work of Kolomeisky \emph{et al.\ }, the authors predicted the correct
functional form of the coupling parameter, but they also needed to correct
their numerical factor $C$ in order to agree with the exact
results~\cite{Kolomeisky1992a}, as we do.}.

\begin{figure}
\center{
\psfrag{xlabel}{$x$ (units of $\ell_{\rm trap}$)}
\psfrag{ylabel}[Bc]{\hspace{2cm}$n(x)$ (units of $\ell_{\rm trap}^{-1}$)}
\epsfig{file=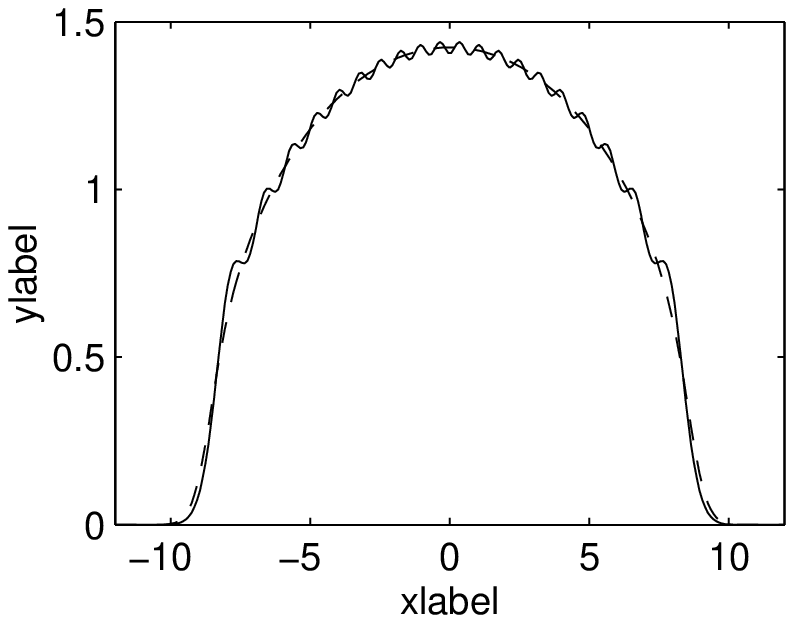,width=8cm}
\Caption{(a) Comparison of ground state density for Tonks gas (solid) and density calculated
from Eq.~(\ref{eq:tonksGPE}) with $|\psi|^4\psi$ non-linearity (dashed),
both with $\mu=20\hbar\omega$ and $N=20$.
\label{fig:chap6_tonksvspsi}}}
\end{figure}

In Fig.~\ref{fig:chap6_tonksvspsi}a we present a comparison of the density
$N|\psi|^2$ obtained from the solution of the GPE of
Eq.~(\ref{eq:tonksGPE}) with the exact result from the Bose-Fermi mapping
theorem calculated by a summation of the first $N$ single-particle states of
the trap.  It can be seen that the GPE solution is in very good agreement with
the exact result, and differs only in the absence of the oscillations seen in
the exact result.  These oscillations rapidly average out as the number of
atoms is increased, and for a moderately large number the difference between
the GPE results and the exact results will be negligible.

\subsection{The delta-function potential}
\label{sec:1DGPEdelta}

We now turn to a discussion of the Lieb-Liniger gas.  We are interested here in
describing the gas from the non-interacting limit right through to the
Tonks-Girardeau limit by means of an appropriate GPE. 

The same argument which we used in deriving the Tonks-Girardeau GPE can again
be used, namely that the coupling parameter should be the many-body T-matrix. 
Again we approximate this by a two-body T-matrix evaluated off the energy shell
at an energy of $-C\mu$, where we use the value of $C=\pi^2/8$ obtained above. 

The two-body T-matrix for a delta-function potential is given
by~\cite{Morgan2002a}
\begin{equation}
\bra{0} T_{\rm 2B}^{\rm LL}(E) \ket{0} = 
\frac{V_0}{1+iV_0 \displaystyle{\left(\frac{m}{2\hbar^2} \right)\sqrt{\frac{\hbar^2}{mE}}}}.
\label{eq:1DdeltaT2B}
\end{equation}
Note that for $V_0 \rightarrow \infty$, this reduces to the
Tonks limit of Eq.~(\ref{eq:offshellT2Btonks}). It follows that our
approximation for the many-body T-matrix in a homogeneous 1D system is
\begin{equation}
g_{\rm LL}(\mu) = \frac{V_0}{1+V_0 \displaystyle{
\sqrt{\frac{2m}{\hbar^2} \frac{1}{\pi^2\mu}}}}.
\label{eq:1DdeltaTMB}
\end{equation}
This energy-dependent form of the coupling parameter can be transformed to a
density-dependent expression by means of the local density approximation, as
outlined above.  This leads to the expression
\begin{eqnarray}
\lefteqn{g_{\rm LL}(x) =}  \label{eq:1Ddeltag}\\
&&{1 \over n(x)}\left( 
\sqrt{ {m\over 2\pi^2\hbar^2}V_0^2 +V_0n(x)} -
\sqrt{{m \over 2\pi^2\hbar^2}}V_0 \right)^2, \nonumber
\end{eqnarray}
which describes an inhomogeneous system.  Note that this expression gives the
correct results for the homogeneous gas in the extreme limits $V_0 \rightarrow
0$ and $V_0 \rightarrow \infty$, giving $\mu = nV_0$ and
Eq.~(\ref{eq:tonkshomogmu}) respectively.

Thus we can now use the above density-dependent expression for the coupling
parameter in Eq.~(\ref{eq:1DGPE}) to obtain a GPE for a trapped 1D gas for any
arbitrary $V_0$.

We use the GPE here to describe the system over the full range of $V_0$,
despite the fact that, as mentioned earlier, a true BEC only exists in a trap
in the weakly-interacting limit~\cite{Petrov2000b}.  Once the interactions
cause sufficient perturbation to the density of states that their discrete
nature is smeared out then no condensate will form in 1D, and in the very
high-$V_0$ (Tonks) limit it is known that there is no condensate.  We can still
use the GPE in the high-$V_0$ regime however, as our previous results have
shown, provided that we now reinterpret the ``wave function'' as simply the
square root of the density $\psi(x) = \sqrt{n(x)}$, rather than the coherent
part of the field operator.

\subsubsection{The Homogeneous Case}

\begin{figure}
\center{
\psfrag{xlabel}{$\gamma$ (dimensionless)}
\psfrag{ylabel}{\hspace{-1cm}$\mu$ (units of $\hbar^2n^2/m$)}
\epsfig{file=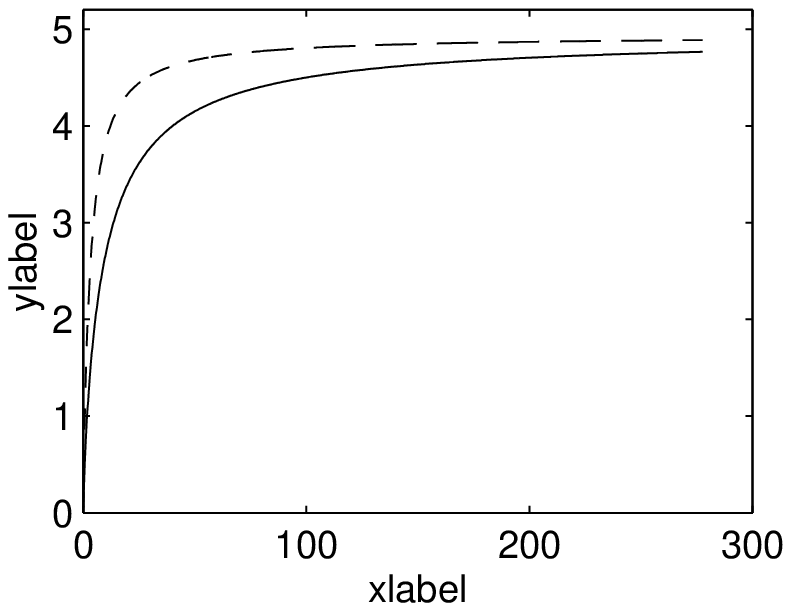,width=8cm}
\Caption{The chemical potential as a function of the parameter $\gamma$ for the
homogeneous Lieb-Liniger gas.  The dashed line shows the exact solution from the
Lieb-Liniger equations~\cite{Lieb1963a}, while the solid line shows the
prediction of the GPE with the coupling parameter given in
Eq.~(\ref{eq:1DdeltaTMB}).  Note that in the extremely high $\gamma$ limit
(not shown)
both lines have the same asymptotic value.
\label{fig:LLandGPEmu}}}
\end{figure}

In the homogeneous case we can compare the results of the GPE with the
interaction strength given by Eq.~(\ref{eq:1Ddeltag}) to the exact results for
a Lieb-Liniger gas.  In Fig.~\ref{fig:LLandGPEmu} we compare the chemical
potentials resulting from the GPE with the exact result of
Eq.~(\ref{eq:tonkshomogmu}).  The results from the GPE show the correct
behaviour in the two extreme limits of $\gamma$, and the overall agreement is
quite good.  However, the chemical potential of the exact results can be seen
to rise more steeply in the intermediate region.  We note that the results have
been obtained using a value of $C$ which strictly applies only in the large
$\gamma$ limit.  A larger value of $C$ gives a curve which agrees very well
with the exact results in the weakly-interacting regime, but predicts the wrong
value in the large $\gamma$ limit.  It seems possible therefore that the value
of $C$ should actually be dependent on $\gamma$, but a detailed investigation
of this is outside the scope of this paper. 

\subsubsection{The Trapped Case}

We now turn to the case of a trapped gas, where we will parameterise our
results both by the interaction strength $V_0$ and by the dimensionless
parameter $\gamma_{\rm trap} = V_0m/\hbar^2\tilde{n}_0$ where $\tilde{n}_0$ is
the peak density of the system (analogous to the parameter used by Lieb and
Liniger).

We expect that in the low-$V_0$ (small $\gamma_{\rm trap}$) limit the chemical
potential will vary as 
\begin{equation}
\mu \propto V_0^{2/3},
\end{equation}
which is the Thomas-Fermi result obtained from the GPE with a cubic
non-linearity (this is not quite the correct prediction in the extremely
low-$V_0$ limit where the Thomas-Fermi approximation is invalid, but it should
describe the majority of the low-$V_0$ regime adequately) .  In the opposite
high-$V_0$ limit, the Bose-Fermi mapping theorem for the trapped
Tonks-Girardeau gas led to Eq.~(\ref{eq:tonksmu}) which is completely
independent of $V_0$.  Note that our derivation of the GPE in
Eq.~(\ref{eq:1DGPE}) neglected terms of order $1/N_0$ since it was derived
in the high-$N_0$ limit.  For this reason we should expect only to reach an
asymptotic chemical potential of $\mu = N_0\hbar\omega$, which is approximately
equal to Eq.~(\ref{eq:tonksmu}) in this limit.

\begin{figure}
\psfrag{chap6deltapsisxlabel}{\hspace{1cm}$x$ (units of $\ell_{\rm trap}$)}
\psfrag{chap6deltapsisylabel}{$\psi(x)$ (units of $\ell_{\rm trap}^{-1/2}$)}
\center{
\epsfig{file=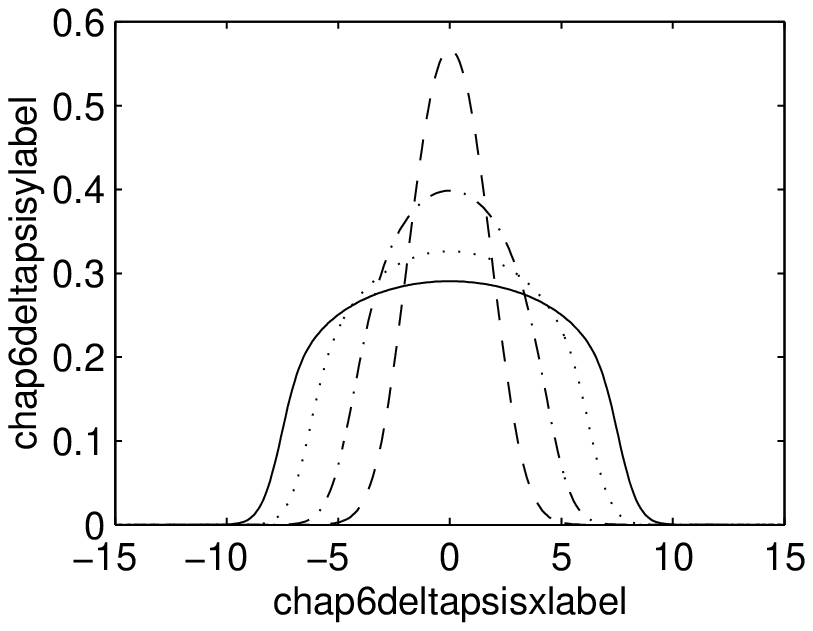,width=8cm}
\Caption{Solutions to the GPE for a 1D gas of $15$ atoms interacting via a
delta-function potential.  The dashed curve corresponds to the
weakly-interacting limit with $V_0 = 0.2\hbar\omega\ell_{\rm trap}$.  The
dot-dash and dotted curves correspond to $V_0 = 3\hbar\omega\ell_{\rm trap}$
and $V_0 = 17\hbar\omega\ell_{\rm trap}$ respectively, while the solid line is
close to the impenetrable gas limit with $V_0 = 250\hbar\omega\ell_{\rm
trap}$.  These values correspond to $\gamma_{\rm trap}=0.02$, $0.63$, $5.3$ and
$99$ respectively. \label{fig:chap6_deltapsis}}}
\end{figure}

\begin{figure}
\psfrag{xlabel1}[cc]{\hspace{1.5cm}$V_0$ (units of $\hbar\omega\ell_{\rm trap}$)}
\psfrag{xlabel2}[Bc]{\hspace{1.5cm}$\gamma_{\rm trap}$ (dimensionless)}
\psfrag{ylabel}[Bc]{\hspace{1.5cm}$\mu$ (units of $\hbar\omega$)}
\center{
\epsfig{file=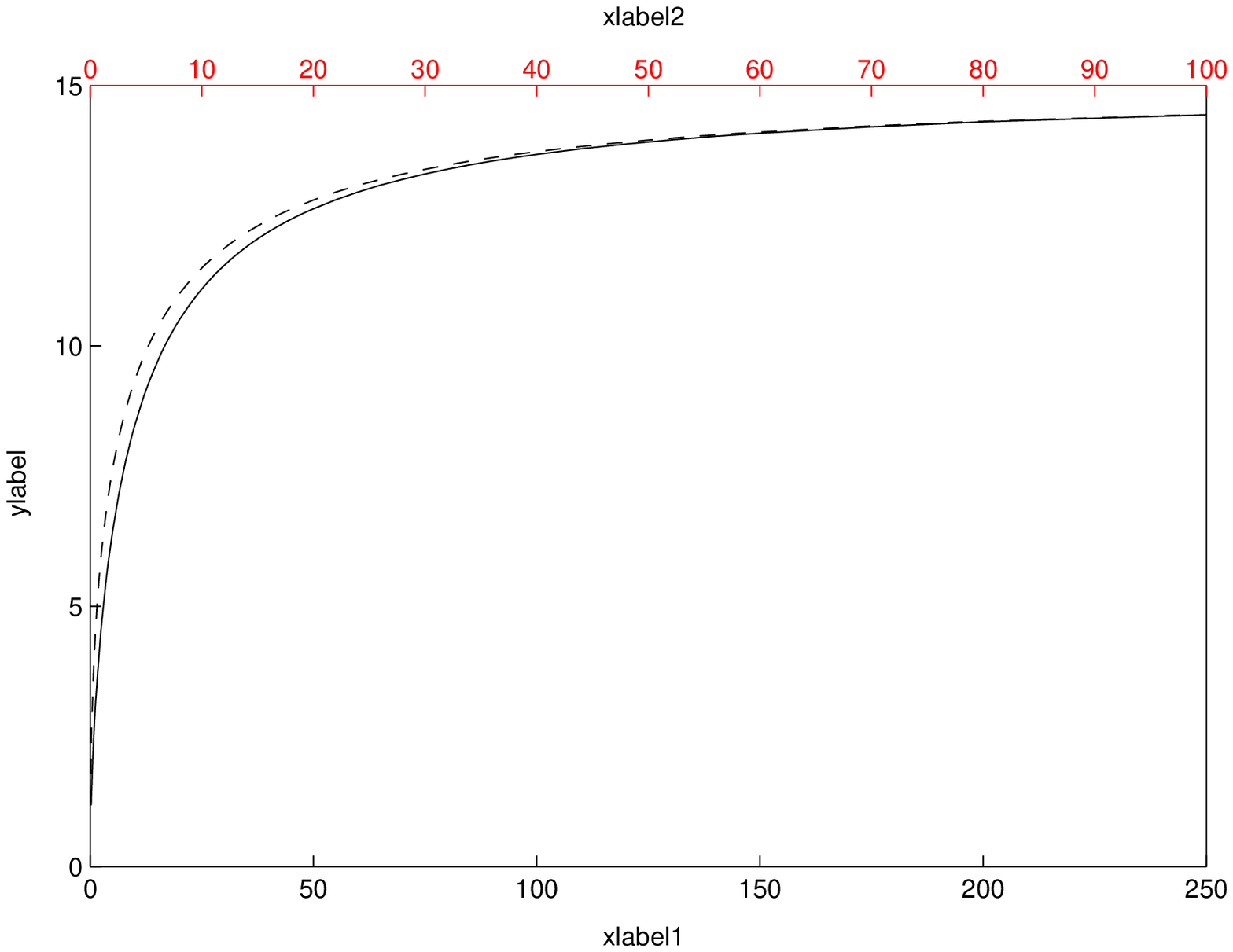,width=8cm}
\Caption{The chemical potential for a gas of $15$ atoms interacting via a
delta-function potential as a function of the potential strength (solid line)
and the parameter $\gamma_{\rm trap}$ (dashed line and upper axis). 
\label{fig:chap6_deltamus}}}
\end{figure}

Figure~\ref{fig:chap6_deltapsis} shows sample solutions to the GPE for a range
of $V_0$ for $N_0=15$.  In the weakly-interacting limit the wave function can
clearly be seen to be a gaussian, as expected.  However as the interactions are
increased the wave function broadens until in the high-$V_0$ limit we have the
shape discussed earlier in the hard-core case.  Figure~\ref{fig:chap6_deltamus}
shows the increase in the chemical potential over the same range.  As expected
from the exact results for the hard-core gas, in the high-$V_0$ limit the
chemical potential becomes almost independent of the interaction strength.  For
the data shown the chemical potential does not appear to quite reach the limit
$\mu = N_0\hbar\omega$ expected from the exact results. However, in a
calculation with the extreme value of $V_0 = 10^4 \hbar\omega\ell_{\rm trap}$
we obtained the result $\mu = 15.00\hbar\omega$ as expected.  This agrees with
the behaviour found by Lieb and Liniger in the homogeneous case, who found that
the results reached the asymptotic limit very slowly (e.g. for $\gamma \approx
4$ the chemical potential had only reached $\sim 0.9$ of its asymptotic
value~\cite{Lieb1963a}).

We have therefore presented a form of the GPE in which the coupling parameter
is an approximate many-body T-matrix given by the off-shell two-body T-matrix.
We have shown that it agrees with the exact results for the Lieb-Liniger gas in
the weakly-interacting and strongly-interacting limits, and it provides an
interpolation between these limits in the intermediate regime.  A drawback of
using this GPE is that we lose any information about the coherences involved in
the system, since we reinterpret the equation simply in terms of the density in
the high-$V_0$ limit.  The concept of using the many-body T-matrix in the GPE
stemmed initially from higher order theories in which it appears naturally via
the anomalous average (defined below).  We therefore turn, in
section~\ref{sec:highorder1D}, to an exploration of these theories in 1D and
see how the changes in the interaction strength arise naturally.

\subsubsection{Quasi-1D gases}

Although we have been dealing with the case of a model delta-function potential
gas, the results can to some extent be mapped onto those for a quasi-1D
system.  In practice, a 1D system could be created by squeezing a 3D trapped
gas condensate in two dimensions.  As the confinement is increased, the system
progresses from being 3D to being 1D, via an intermediate regime we call
quasi-1D for the purposes of this paper.  In this intermediate regime, the
dynamics of the system as a whole are frozen out in two dimensions, but the
scattering between particles is not yet genuinely one-dimensional. 

In the 3D regime the system is described by a GPE with a $|\psi|^2\psi$
nonlinearity, as for the low $V_0$ Lieb-Liniger gas.  However, in the strict 1D
limit the atoms will be unable to scatter past one another, and will form a
Tonks-Girardeau gas.  In this limit the appropriate GPE will be that of
Eq.~(\ref{eq:tonksGPE}).  The progression from a 3D gas to a 1D gas
therefore mirrors in many respects the progression which occurs if the potential
strength $V_0$ is increased for the Lieb-Liniger gas.  The 3D to 1D progression
has been investigated in more detail in references~\cite{Olshanii1998a,Dunjko2001a,Bergeman2002a}.

\section{Higher order theories in 1D}
\label{sec:highorder1D}

In the previous section we used the GPE in conjunction with an approximate
many-body T-matrix in order to describe the Lieb-Liniger gas.  In this section
we examine the suitability of higher-order theories to describe this gas. 
Ideally we would like a theory which makes the transition between the weakly
interacting and strongly-interacting regimes smoothly, and which accurately
describes the coherence of the system as well as the densities.  There are
several higher order theories which are used to good effect to describe 3D
condensates, and in this section we will apply some of these to the 1D trapped
gas case. 

Since we can always choose a value of $V_0$ to be sufficiently weak that a 
condensate does exist in a trapped zero-temperature 1D gas, there will always
be a regime in which the 3D theories can be applied.  As we increase $V_0$ the
condensate is replaced by a quasicondensate with only short range phase
coherence, and this will lead to problems in the higher order theories. 
However, we have described in the previous section the behaviour that we expect
in the high-$V_0$ limit. It is therefore of interest to apply the higher order
theories to see how and why they fail in this limit, and to gain insight into
any improvements needed to describe the 1D gas adequately.

As we have seen, the many-body effects on interactions are fundamental in
understanding the behaviour of low dimensional BECs.  We must therefore use a
theory in which these effects are accounted for accurately.  Such a theory is
provided by the gapless-Hartree-Fock-Bogoliubov (GHFB)
theory~\cite{Proukakis1998a,Hutchinson2000a}.  In this theory, the condensate
wave function $\psi_0(x)$ is described by the generalised GPE in 1D
\begin{eqnarray}
\mu \psi_0(x)&=&-{\hbar^2 \over 2m}{d^2 \over dx^2}\psi_0(\br) 
+ V_{\rm trap}(x)\psi_0(x)\nonumber \\
&&\mbox{}+ N_0U_{\rm con}(x)|\psi_0(x)|^2\psi_0(x) \nonumber \\
&&\mbox{}+2U_{\rm ex}(x)
\rho(x)\psi_0(x).
\label{eq:c6gengpe}
\end{eqnarray}
Here $\rho(x)$ is the non-condensate density and it is defined, along with the
associated anomalous average $\kappa$ [needed in Eq.~(\ref{eq:TMBc6})], by
\begin{eqnarray}
\rho(x) &=& \sum_{i\neq 0}\left(|u_i(x)|^2+|v_i(x)|^2\right)n_i 
+ |v_i(x)|^2, \label{eq:rho} \\
\kappa(x) &=&\sum_{i\neq 0}u_i(x)v_i^*(x)\left( 2n_i +1 \right).
\label{eq:kappa}
\end{eqnarray}
The Bogoliubov functions $u_i(x)$ and $v_i(x)$ are the solutions of 
\begin{equation}
\begin{array}{lcr}
{\cal L}(x)u_i(x) + {\cal M}(x)
v_i(x) &=& 
\varepsilon_iu_i(x), \\[2mm]
{\cal L}(x)v_i(x) + {\cal M}^*(x)
u_i(x) &=& 
-\varepsilon_iv_i(x),
\end{array}
\end{equation}
solved in a basis orthogonal to $\psi_0(x)$, and with
\begin{eqnarray}
{\cal L}(x)  &=& \hat{H}_{\rm sp} - 
\mu +2N_0U_{\rm con}(x) |\psi_0(x)|^2 \nonumber \\ 
&&\mbox{}+2U_{\rm ex}(x)\rho(x), 
\label{eq:c6L}
\\
{\cal M}(x) &=& N_0U_{\rm con}(x)\psi_0^2(x).
\end{eqnarray}
The quasiparticle population factors $n_i$ are given by the Bose-Einstein distribution function, and for the zero temperature case we are concerned with here these factors vanish.

This theory distinguishes between the interactions which occur between two
condensate atoms (governed by $U_{\rm con}(x)$), and those which occur between
a condensate atom and a non-condensate atom (governed by $U_{\rm ex}$). The
various higher order theories which we will explore in this section can now be
summarised thus:
\begin{description}
\item[BdG]  By setting $U_{\rm ex}= 0$ (i.e. neglecting interactions between
the condensate and the non-condensate), and using $U_{\rm con}(x) = V_0$, we
obtain the familiar GPE and Bogoliubov-de Gennes equations.  
\item[HFB-Popov]  The HFB-Popov theory corresponds to the choice $U_{\rm
con}(\br) = V_0$ and $U_{\rm ex}(\br) = V_0$.  Thus both condensate-condensate
interactions and condensate-non-condensate interactions are included, although
many-body effects on the interaction strength are not.
\item[GHFB1]  The first version of the GHFB theory  is given by setting $U_{\rm
con}(\br) = T_{\rm MB}(\br)$ [defined in Eq.~(\ref{eq:TMBc6})] and $U_{\rm
ex}(\br) = V_0$.  This version of GHFB therefore improves upon the description
of the BdG or HFB-Popov descriptions by including many-body effects on the
scattering of two condensate atoms.
\item[GHFB2]  From the above arguments it is clear that another GHFB theory may
be proposed.  This consists of setting $U_{\rm con}(\br) = T_{\rm MB}(\br)$ and
$U_{\rm ex}(\br) = T_{\rm MB}(\br)$, which upgrades the description of both
condensate-condensate interactions and condensate-non-condensate interactions
to the many-body T-matrix.  
\end{description}

In the GHFB theories, the many-body T-matrix is given
by~\cite{Proukakis1998a,Morgan2000a}
\begin{equation}
T_{\rm MB}(x) = V_0\left(1 + {\kappa(x) \over N_0\psi_0^2(x)} \right),
\label{eq:TMBc6}
\end{equation}
and so we can see that the inclusion of the anomalous average leads to the
inclusion of the many-body T-matrix in the theories.

Note that we may use the bare potential $V_0$ throughout these equations, 
since we consider the \emph{model} Lieb-Liniger system in which the interaction
potential truly is a delta function.  Furthermore, in 1D there is an absence of
the ultra-violet divergences which occur in other dimensions when a contact
potential is used.  Since the above equations are always solved
self-consistently, we still generate the many-body T-matrix (where used) to all
orders from the bare potential~\cite{Proukakis1998a}.

\subsection{Condensate and quasi-condensate properties}

We start our simulations in the weakly-interacting limit with a small value of
$V_0$.  In this limit Lieb and Liniger~\cite{Lieb1963a} found that the
Bogoliubov theory gave a good description of the homogeneous system, and so we
expect our model should also be valid.  We solve the above equations
self-consistently to obtain the wave function, chemical potential, many-body
T-matrix,  $\rho(x)$, and $\kappa(x)$.  We then ramp up the value of $V_0$,
using the results from the previous iteration as our starting point.  By this
method we have managed to solve the equations from the weakly-interacting limit
into the strongly-interacting limit where the system should become a
Tonks-Girardeau gas.  Again, we expect the chemical potential to change
initially as $\mu \propto V_0^{2/3}$ at low $V_0$, before reaching a plateau
at  $\mu = N_{\rm tot}\hbar\omega$ in the strongly-interacting limit.

We present the results from the different forms of the theory below.  In all
cases we have used $N_0 = 15$, and we start our simulations with $\tilde{V_0} =
V_0/\hbar\omega\ell_{\rm trap} = 0.2$ which is sufficiently weak an interaction
that a solution is obtained if $\rho$ and $\kappa$ are initially zero.

\begin{figure}
\psfrag{xlabela}{\hspace{0.5cm}$\tilde{V}_0$}
\psfrag{ylabela}{$\mu/\hbar\omega$}
\psfrag{xlabelb}{\hspace{0.5cm}$\tilde{V}_0$}
\psfrag{ylabelb}{${\mu/N_{\rm tot}\hbar\omega}$}
\center{
\epsfig{file=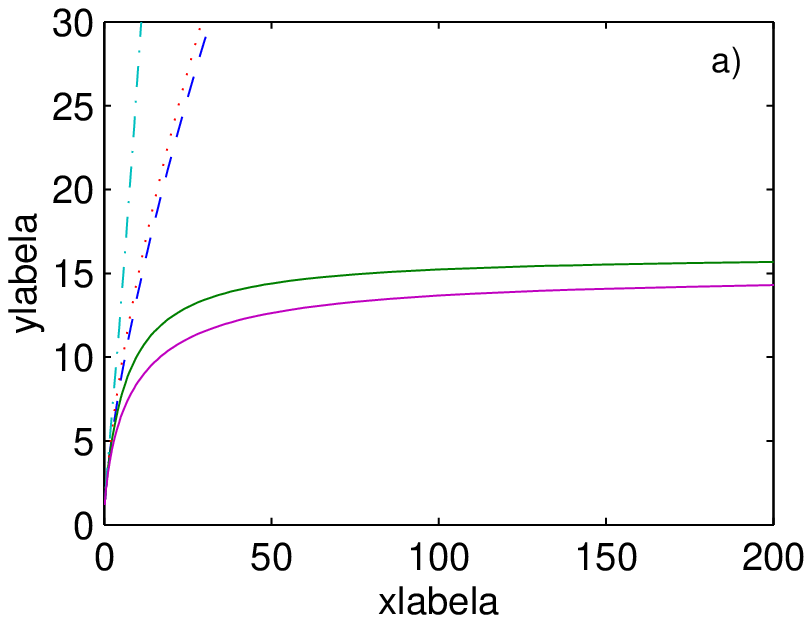,width=8cm}
\epsfig{file=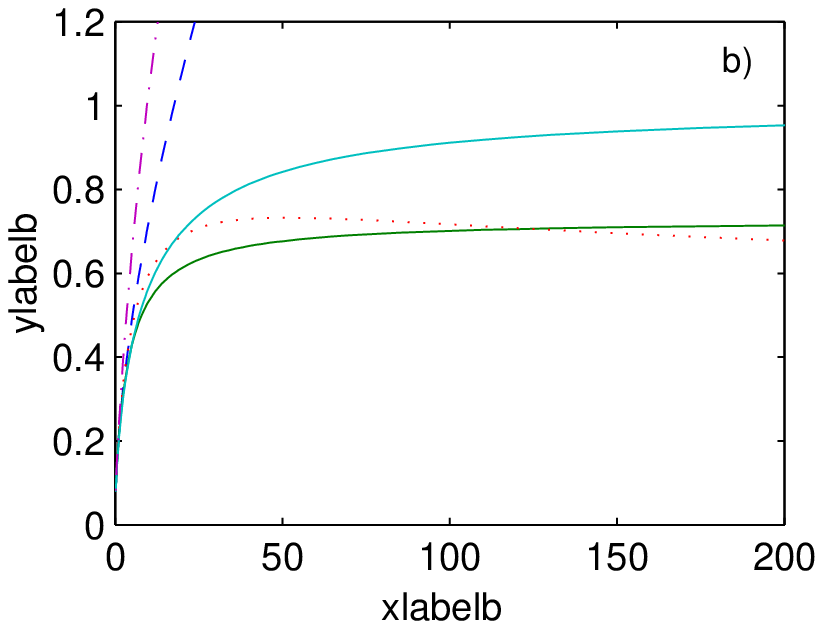,width=8cm}
\Caption{a) The chemical potential as a function of the interaction potential
$\tilde{V}_0= V_0/\hbar\omega\ell_{\rm trap}$ as calculated by the various
different models. The solid lines represent the results from the GHFB2 theory
(upper line) and the GPE results of the previous section (lower line).  The
other results correspond to the GHFB1 theory (dashed), HFB-Popov theory
(dash-dot) and the BdG equations (dotted). b) The chemical potential per atom
in the system as a function of $V_0$.  The exact results predict that
$\mu/N_{\rm tot}\hbar\omega= 1$ in the high $V_0$ limit.  The broken lines
correspond to the same curves as in part (a), while the solid lines swap - the
\emph{lower} curve representing the GHFB2 results in this figure. The parameter
$\gamma_{\rm trap}$ is not plotted, however for comparison purposes
$\tilde{V}_0 =200$ roughly corresponds to $\gamma_{\rm trap} \approx
80$.\label{fig:chap6_muvsV}}}
\end{figure}

In Fig.~\ref{fig:chap6_muvsV}a the chemical potential has been plotted as a
function of the potential strength $V_0$, for the various different models,
including the GPE results from the previous section. 
Figure~\ref{fig:chap6_muvsV}b also shows $\mu/N_{\rm tot}$ the chemical
potential per atom, in order to compare more easily to the expected high-$V_0$
limit $\mu = N_{\rm tot}\hbar\omega$.  Note that $N_{\rm tot}$ is the total
population of the system, consisting of both the condensate and non-condensate
populations.  In the low-$V_0$ limit it can be seen that all of the models
provide very similar results, as should be expected since the gas is
weakly-interacting in this limit.   As $V_0$ is increased, however, the
different models show quite different behaviour.

In terms of the chemical potential, neither the BdG nor the HFB-Popov models
give the required behaviour, instead $\mu$ increases rapidly as $V_0$ is
increased.  In neither model does the graph show a plateau in the high-$V_0$
limit, and the chemical potential greatly exceeds the Tonks-Girardeau limit
$\mu = N_{\rm tot}\hbar\omega$ in the HFB-Popov case.  Instead the chemical
potential roughly follows the $V_0^{2/3}$ behaviour for all $V_0$.  The
differences between these two models is seen in Fig.~\ref{fig:chap6_muvsV}b,
in which the results differ due to the different total populations.  In the BdG
case the non-condensate population increases very rapidly, such that in the
very high-$V_0$  limit $\mu/N_{\rm tot}$ is actually decreasing.  In the
HFB-Popov model, the interactions between the condensate and non-condensate
slow the growth of the non-condensate population, but the chemical potential
still rises unchecked.

The disagreement of these models with the exact results is not at all
surprising.  In both models the anomalous average is completely neglected, and
so the many-body T-matrix is not introduced at all.  We have argued here, and
in earlier papers~\cite{Lee2002a,Lee2002b}, about the importance of many-body
effects in interactions in low-dimensions, and so we expect that neglecting
these effects will lead to an inadequate theory.  The BdG and HFB-Popov results
presented here are similar to those obtained in the homogeneous gas by Lieb and
Liniger~\cite{Lieb1963a}, who found that the standard Bogoliubov method failed
at high interaction strengths, when $\gamma \gtrsim 2$.  In our results we also
see quite good agreement with the GPE of the previous section for $\gamma_{\rm
trap} \lesssim 1$ and the differences between the theories become much greater
when $\gamma_{\rm trap}$ is increased above this point \footnote{$\gamma_{\rm
trap}$ has not been plotted on Fig.~\ref{fig:chap6_muvsV} since it varies
slightly differently for each curve.  However a good idea of its approximate
value may be obtained by comparing with Fig.~\ref{fig:chap6_deltamus}.}.  Note
that $\gamma_{\rm trap} \sim 2$ corresponds to $\tilde{V}_0 \sim 10$ in the
figure, which is roughly the point at which the curves separate. 

The anomalous average, and therefore the many-body T-matrix, can first be
included by solving the GHFB1 equations.  Figure~\ref{fig:chap6_muvsV} shows
that there is little difference between these results and those of the
HFB-Popov theory.  The chemical potential still easily exceeds the limit of the
Tonks-Girardeau gas, and so the GHFB1 theory also fails rapidly as $V_0$
increases.

\begin{figure}
\center{
\psfrag{xlabel}[Bc]{\hspace{2cm}$x$ (units of $\ell_{\rm trap}$)}
\psfrag{ylabel}[Bc]{\hspace{1.5cm}$T_{\rm MB}(x)/N_0|\psi(x)|^2$ (arb. units)}
\epsfig{file=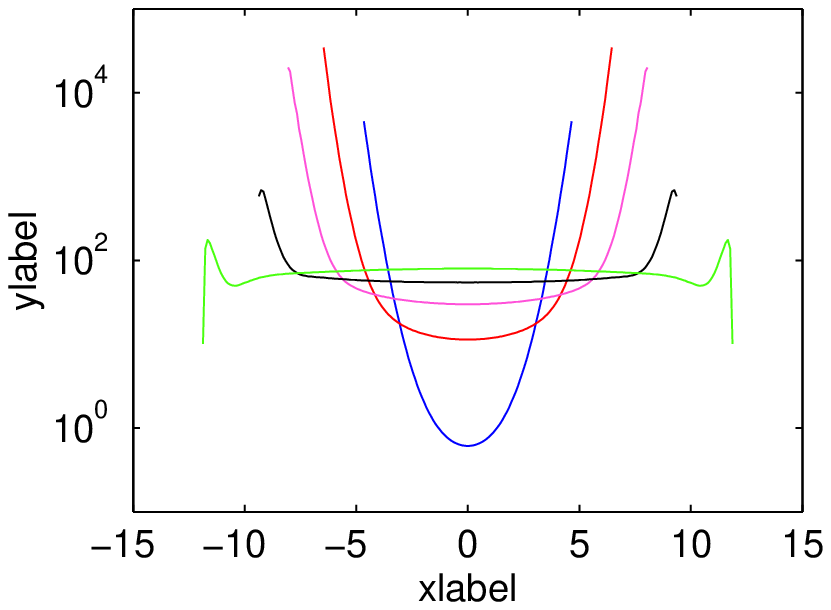,width=8cm}
\Caption{$T_{\rm MB}(x)/N_0|\psi(x)|^2$ for the many-body T-matrix calculated
from the GHFB1 theory for various values of $V_0$.  Note that the ratio is only
plotted for the range over which the condensate density is significant.  The
values of $\tilde{V}_0$ used were $0.2$, $2.4$, $6.5$, $15$ and $44$, which
correspond to the narrowest to the widest curves respectively.  These values
correspond to $\gamma_{\rm trap} = 0.02$, $0.46$, $1.6$, $4.3$ and $16.5$.
\label{fig:chap6_GHFB1Tmatrix}}}
\end{figure}

However, the GHFB1 results do allow us to examine the form of the many-body
T-matrix.  In Fig.~\ref{fig:chap6_GHFB1Tmatrix} the spatial dependence of the
many-body T-matrix is plotted as $V_0$ is increased.  In fact, the value being
plotted is the T-matrix divided by the condensate density $|\psi(x)|^2$.  From
the previous section we expect that the T-matrix becomes proportional to the
condensate density as $V_0 \rightarrow \infty$, as in
Eq.~(\ref{eq:1Dspatialgtonks}).  Figure~\ref{fig:chap6_GHFB1Tmatrix} shows
that as the interaction strength is increased this becomes an increasingly good
approximation.  The very good agreement with the form expected for large $V_0$
is further evidence that the coupling parameter $g_{\rm LL}(x)$ derived earlier
is accurate in this limit. This verifies our approximation for the many-body
T-matrix in terms of the off-shell two-body T-matrix numerically. 

That the GHFB1 theory fails in the high-$V_0$ limit is not particularly
surprising.  In the low-$V_0$ limit almost all the atoms are in the
condensate.  As the interaction strength is increased, however, we reach the
regime of the quasicondensate discussed by Petrov \emph{et al.\
}\cite{Petrov2000b}. As $V_0$ is increased the number of atoms in excited
states (i.e. $\rho$) increases, and can become a sizable proportion of the
total number of atoms in the system.  It therefore becomes important to
describe the interactions between these atoms accurately.  

The description of interactions between excited atoms is more complicated than
those between condensate atoms.  However, two limits may be investigated
relatively straightforwardly.  If the majority of the collisions occur at a
high energy then the correct description would be given by the two-body
T-matrix (since the many-body T-matrix tends towards the two-body T-matrix in
this limit).  On the other hand, if the majority of the collisions are at low
energy, then it is more appropriate to describe them by the same many-body
T-matrix as used to treat the condensate-condensate interactions.  In 2D and 3D
it might be expected that the large density of states at high energies should
mean that the majority of collisions are in this regime. However, in 1D the
density of states is independent of energy in a trap.  In the zero-temperature
limit in which we are working it therefore seems probable that the appropriate
description of the interactions involving non-condensate atoms is the same
many-body T-matrix used to describe the condensate interactions.

The many-body T-matrix is used to describe non-condensate interactions in the
GHFB2 version of the theory.  It can be seen from Fig.~\ref{fig:chap6_muvsV}
that the results using this theory are quite different from those of the GHFB1
theory.  In the low-$V_0$ limit the results are similar, but as the
interactions are increased the chemical potential becomes independent of $V_0$,
as we expect in the Tonks-Girardeau limit.  The figure clearly shows that it
reaches an asymptotic value of $\sim 0.7 N_{\rm tot}\hbar\omega$.  This is
lower than the expected value of $N_{\rm tot}\hbar\omega$, but nonetheless the
results are in qualitative agreement with the behaviour expected for the Tonks
gas.

Again, we can compare the spatial shape of the many-body T-matrix with our
earlier approximation, and the results look essentially the same as those
calculated from the GHFB1 results (Fig.~\ref{fig:chap6_GHFB1Tmatrix}).  The
major change in the GHFB2 theory is therefore to provide a better description
of the condensate-non-condensate interactions rather than any implicit change
in  the form of the many-body T-matrix.

\begin{figure}
\psfrag{xlabel}[cc]{\hspace{1cm}$x$ (units of $\ell_{\rm trap}$)}
\psfrag{ylabel}[Bc]{\hspace{1cm}Total density}
\center{
\epsfig{file=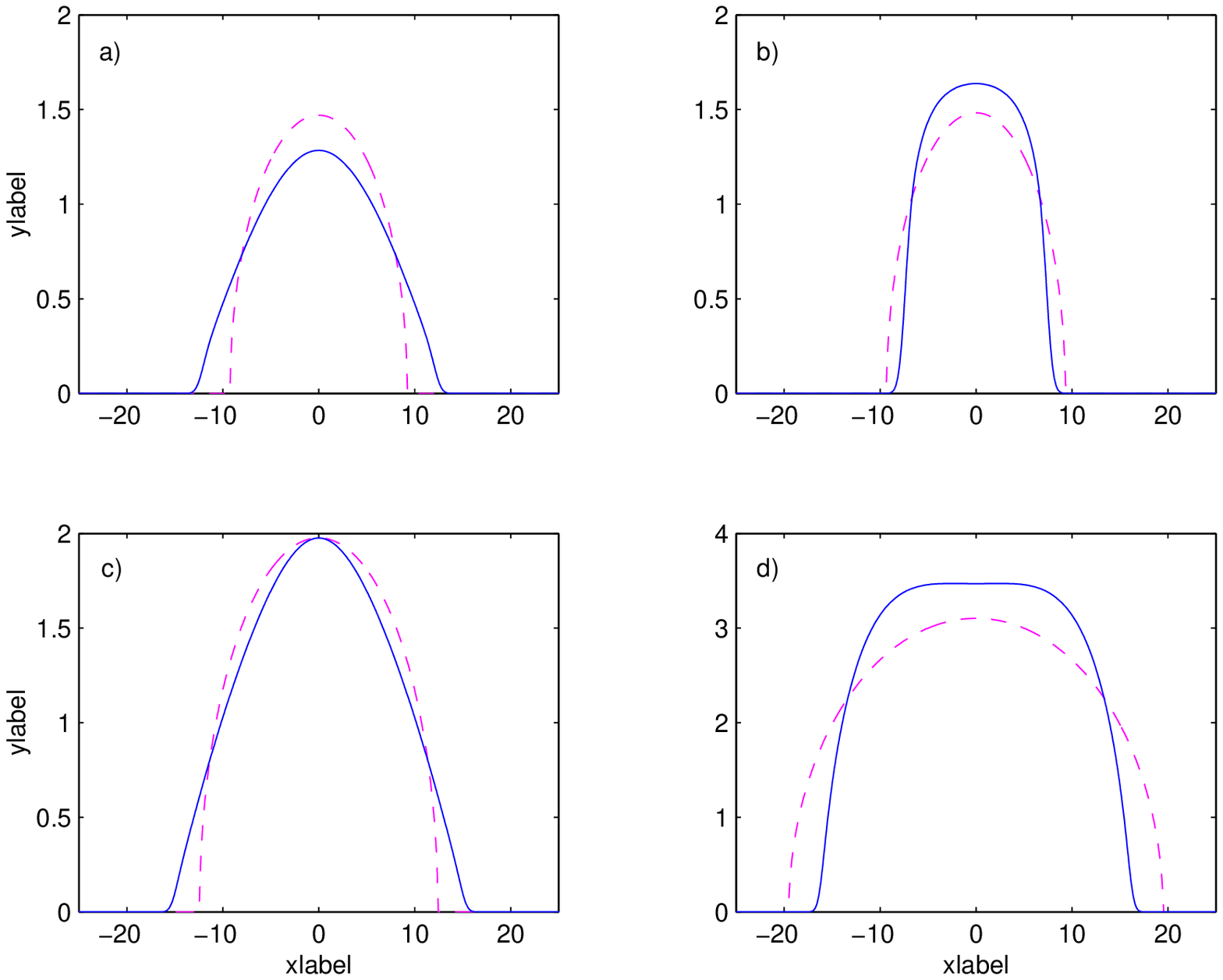,width=8cm}
\Caption{Total density $N_0|\psi(x)|^2 +\rho(x)$ (solid) compared to the
Quintic GPE prediction in the Thomas-Fermi limit (dashed) in the high $V_0$
regime for the different theories. a) GHFB1 results at $\tilde{V}_0 = 50$
($\gamma_{\rm trap} = 20$) (this was as high a potential for which consistent
results were obtained).  b) GHFB2 results at $\tilde{V}_0 = 100$ ($\gamma_{\rm
trap} = 30$). c) HFB-Popov results at $\tilde{V}_0 = 20$ ($\gamma_{\rm trap} =
5$) (again, results for potentials higher than this were not obtained). d) The
BdG equation results at  $\tilde{V}_0 = 100$ ($\gamma_{\rm trap} = 15$).
\label{fig:chap6_densities}}}
\end{figure}

We can also compare the predicted density distribution in the high-$V_0$ limit
with the solutions of the quintic GPE which was shown to give good agreement
with the Tonks-Girardeau results in section~\ref{sec:1DGPE}.  In
Fig.~\ref{fig:chap6_densities} we compare the total densities
$[N_0|\psi(x)|^2 +\rho(x)]$ for the different models with the Thomas-Fermi
prediction from Eq.~(\ref{eq:quinticGPETF}).  It can easily be seen that
the best agreement is given by the GHFB2 theory, while the BdG, HFB-Popov and
GHFB1 results all give entirely the wrong shape.  It must be said however, that
the GHFB2 results are still not in good agreement with the predicted density.

Thus the GHFB2 theory interpreted in the normal manner fails to quantitatively
describe the 1D system in the impenetrable limit. This is not surprising since
it assumes the existence of a condensate while we know that a condensate does
not exist in this limit.  Nevertheless, the GHFB2 theory does give
qualitatively the correct behaviour for the chemical potential and density in
the high-$V_0$ limit, which is a significant improvement on the other forms of
mean-field theory used for 3D condensates.  

The failure of the GHFB1 theory and the relative success of the GHFB2 theory
indicate that the description of the non-condensate is the key to solving the
1D problem in the strongly-interacting limit.  The GHFB1 model fails at high
$V_0$ since an increase in $V_0$ leads to an increase in $\rho$.  The $\rho$
term obviously becomes far too dominant in GHFB1 and therefore needs to be
suppressed by some means.  In GHFB2 the many-body T-matrix is introduced to
describe interactions involving $\rho$, and since $T_{\rm MB}/V_0$ decreases
with increasing $V_0$ this achieves the required suppression.  As confirmation
of this argument, the GHFB1 results were rerun with $\rho$ artificially set to
zero.  With this adjustment, the GHFB1 model gave results which looked
qualitatively similar to the GHFB2 results, with the chemical potential
reaching a plateau in the high-$V_0$ limit.

\subsection{Quasiparticle spectrum}

Having seen qualitative agreement between the GHFB2 theory and the expected
behaviour of the chemical potential and the density, we turn now to investigate
the quasiparticle spectrum.  In the high-energy limit the quasiparticles are
predicted to be single-particle like, and therefore for a 1D harmonic potential
the energy will rise linearly with the quasiparticle quantum number.  However,
for the low-energy quasiparticle states (with energies less than $\sim \mu$),
we can use the hydrodynamic theory to obtain an expression for the
quasiparticle energies $\varepsilon_n$ in the Thomas-Fermi
limit~\cite{Stringari1996a}.  Using the same method as
in~\cite{Stringari1996a}, we find that the low-energy quasiparticle spectrum
for a BEC in 1D described by a cubic GPE is
\begin{equation}
\varepsilon_n = \hbar\omega\sqrt{{n(n+1)\over 2}}, \label{eq:qpEscubic}
\end{equation}
where $n = 1,2,...$ and $n=1$ is the Kohn mode.  However, we have shown earlier
that in the impenetrable limit the system is described instead by a quintic
GPE.  For such a GPE the same hydrodynamic approach instead
predicts~\cite{Minguzzi2001a} 
\begin{equation}
\varepsilon_n = n\hbar\omega,  \label{eq:qpEsquintic}
\end{equation}
which is obviously what one would expect for non-interacting fermions, and so
we would expect this relationship from a correct theory in the high-$V_0$
limit.

\begin{figure}
\psfrag{xlabel}[cl]{$\tilde{V}_0$}
\psfrag{ylabela}{$\varepsilon_1/\hbar\omega$}
\psfrag{ylabelb}{$\varepsilon_2/\hbar\omega$}
\psfrag{ylabelc}{$\varepsilon_3/\hbar\omega$}
\psfrag{ylabeld}{$\varepsilon_4/\hbar\omega$}
\center{
\epsfig{file = 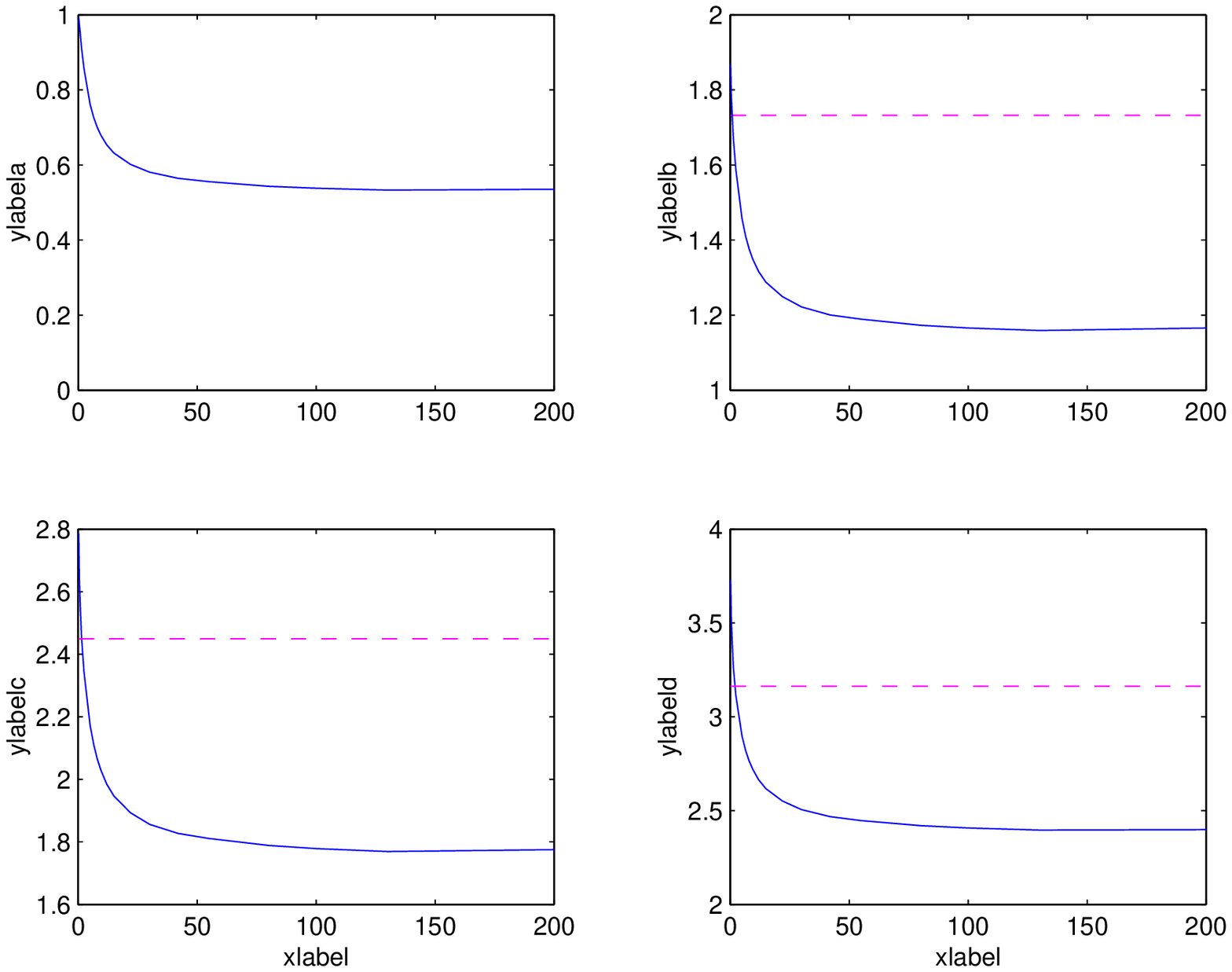,width=8cm}
\Caption{The energies of the first four quasiparticle states as a function of
the interaction potential $\tilde{V}_0$, as predicted by GHFB2 theory.  The
dashed lines indicate the value expected from hydrodynamic theory by the
linearisation of the cubic GPE (for the Kohn mode [$\varepsilon_1$] the
expected value is $1\hbar\omega$).  The correct values in the high-$V_0$ limit
are $\hbar\omega$, $2\hbar\omega$, $3\hbar\omega$, and $4\hbar\omega$ (from left
to right).
\label{fig:chap6_GHFB2energies}}}
\end{figure}

Figure~\ref{fig:chap6_GHFB2energies} shows the GHFB2 results for the energies
of the first four quasiparticle modes as a function of the interaction
strength.  In the weakly-interacting limit the energies are all close to the
prediction of Eq.~(\ref{eq:qpEscubic}), as expected.  However, as the
interaction strength is increased, the quasiparticle energies do not go over to
the limit of Eq.~(\ref{eq:qpEsquintic}).  Instead they all fall to around
60\% of that value and, even in the high-$V_0$ limit, the relationship between
the different energies scales more like Eq.~(\ref{eq:qpEscubic}) than
Eq.~(\ref{eq:qpEsquintic}).  The conclusion which must be drawn is that
the GHFB2 theory does not predict the quasiparticle energies at all accurately
in the high-$V_0$ limit.  This can be seen particularly strongly in the case of
the Kohn mode ($\varepsilon_1$), which should be equal to $\hbar\omega$ at all
interaction potentials since it simply corresponds to a centre-of-mass
oscillation of the entire system in the harmonic potential.

The reason that GHFB2 does not predict the quasiparticle spectrum well is
because it does not include any dynamics of the non-condensate.  In the
strongly-interacting limit a large proportion of the atoms in the system are in
the non-condensate, which is assumed to be completely static in all of the
theories discussed here.  Since the Kohn mode corresponds to the motion of all
of the atoms together, it is not surprising that a model which does not include
the dynamics of a large proportion of those atoms predicts the wrong frequency.

To see this in more detail,  all of the theories discussed here can
be derived by starting from the generalised time-dependent GPE
\begin{eqnarray}
\lefteqn{i\hbar {\partial \Psi(x,t) \over \partial t} = } \nonumber \\
&&-{\hbar^2 \over 2m}{\partial^2
\over \partial x^2}\Psi(x,t) +N_0U_{\rm con}(x)|\Psi(x,t)|^2\Psi(x,t) \nonumber \\
&&\mbox{}+V_{\rm trap}(x)\Psi(x,t)  
+ 2U_{\rm ex}(x)\rho(x)
\Psi(x,t), \label{eq:TDGPE}
\end{eqnarray}
and expanding $\Psi(x,t)$ in terms of the unperturbed ground-state $\psi(x)$
and time-dependent fluctuations $\Delta\psi(x,t)$, and linearising (see, for
example,~\cite{Ruprecht1996a}).  The trouble occurs when it is recognised that
in 1D the appropriate coupling parameter $U_{\rm con}(x)$ is the many-body
T-matrix, as argued earlier.  In the high-$V_0$ limit, the many-body T-matrix
is proportional to $N_0|\psi(x)|^2$ in the static case as we have shown. 
Applying the local density approximation, the time-dependent form of the
many-body T-matrix should then be $N_0|\Psi(x,t)|^2$ (assuming that the
time-scale for interparticle interactions is much quicker than that for
condensate dynamics).  In the linearisation procedure, one should therefore
also expand the many-body T-matrix as
\begin{equation}
T_{\rm MB}(x) \propto |\psi(x)+\Delta\psi(x,t)|^2.
\end{equation}
Making this replacement for $U_{\rm con}(x)$ in Eq.~(\ref{eq:TDGPE}) while
setting $\rho=0$ leads directly to the quintic GPE discussed earlier and hence
to the result of Eq.~(\ref{eq:qpEsquintic}).  We see therefore that
including the dynamics of the anomalous average in the GHFB2 theory will lead
to the correct prediction for the excitation spectrum in the high-$V_0$ limit.
For a general value of $V_0$, including the fluctuations of the many-body
T-matrix requires a dynamic treatment of the anomalous average and hence of the
non-condensate generally.

\subsection{Coherences}

It is of interest to calculate the coherence functions for the Lieb-Liniger gas
from the GHFB2 results.  We have argued that we are dealing with a gas in which
a condensate may form in the low-$V_0$ limit, but where it disappears when the
interactions are increased.  Since coherence is a major feature of a condensate
we should be able to see a loss of coherence as $V_0$ is increased, and again we can try to relate this to the expected results in the Tonks limit.

\subsubsection{First-order coherence}

The first order coherence function $G^{(1)}(\br,\br')$ is defined in terms of
the quantum field operators by~\cite{Naraschewski1999a}
\begin{equation}
G^{(1)}(\br,\br') = \langle \hat{\Psi}^\dagger(\br)\hat{\Psi}(\br')\rangle,
\end{equation}
and the normalised first order coherence function is therefore
\begin{equation}
g^{(1)}(\br,\br') = {G^{(1)}(\br,\br') \over \sqrt{G^{(1)}(\br,\br)
G^{(1)}(\br',\br')}}.
\end{equation}
Decomposing the field operator in terms of the condensate and its fluctuations
leads to an expression for the coherence function at $T=0$ in terms of the
condensate wave function and Bogoliubov quasiparticle
amplitudes~\cite{Dodd1997c}
\begin{equation}
G^{(1)}(\br,\br') = N_0\psi_0^*(\br)\psi_0(\br') + 
\sum_{i \neq 0}v_i(\br)v_i^*(\br'). \label{eq:G1T0}
\end{equation}

\begin{figure}
\psfrag{xlabel}[cl]{$x/R_{\rm TF}$}
\psfrag{ylabel}[bl]{$g^{(1)}(0,x)$}
\center{
\epsfig{file = 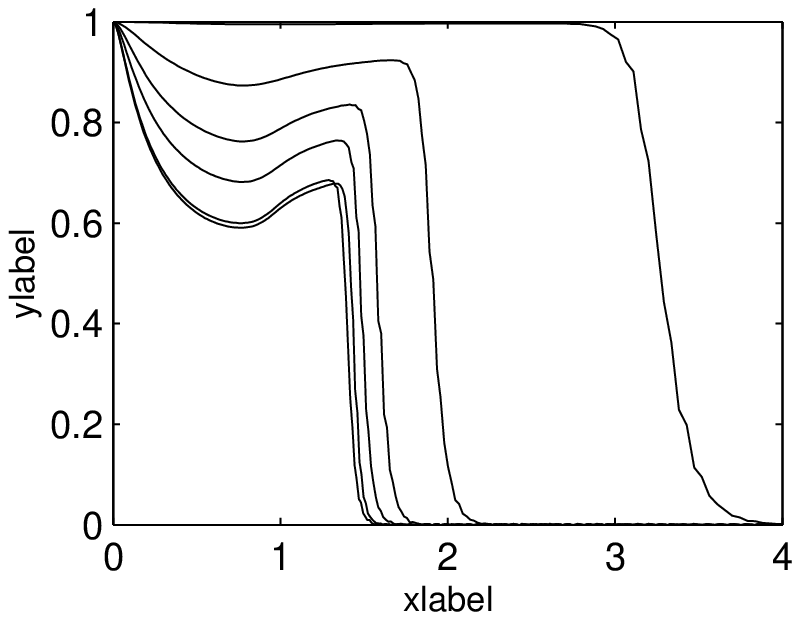,width=8cm}
\Caption{The normalised first order coherence function $g^{(1)}(0,x)$
calculated using the GHFB2 theory for various values of $V_0$.  The solid
curves represent (from top to bottom) the results for $\tilde{V}_0 =
0.2,2.4,6.5,15,100$ and $200$ (corresponding to $\gamma_{\rm trap} =
0.02,0.46,1.6,4.1,31$ and $61$ respectively).  The coherence function is plotted as a function of $x/R_{\rm TF}$ where $R_{\rm TF}$ is the Thomas-Fermi
radius of the condensate. \label{fig:chap6_coherencesfig2}}}
\end{figure}

Figure~\ref{fig:chap6_coherencesfig2} shows the normalised first order
coherence function obtained from the GHFB2 results for various values of
$V_0$.  Note that the figure has been plotted in terms of the distance scale
$\tilde{x} = x/R_{\rm TF}$, where $R_{\rm TF}$ is the Thomas-Fermi radius of
the condensate, in order to take into account the greater spatial extent of a
condensate with stronger interactions.  Figure~\ref{fig:chap6_coherencesfig2}
clearly shows that the system is coherent over its entire length in the
weakly-interacting limit.  However, as the interaction strength is increased
the coherence can be seen to diminish substantially, and the width of the
central maximum decreases as well.  A diminishing range of coherence is
consistent with the absence of a true condensate in the strongly-interacting
regime.  However, the rate at which the coherence calculated here decreases is
surprisingly slow.  Gangardt and Shlyapnikov~\cite{Gangardt2002a} have
presented results for the coherence function obtained from hydrodynamic theory
applied to a 1D gas, and they see a much greater fall in the range of the
coherence length - dropping by almost an order of magnitude as $\gamma_{\rm
trap}$ decreases from $0.1$ to $10$. However, their results were obtained for a
system of $10^4$ atoms, many more than we deal with here.  Also, even for the
largest value of $V_0$ plotted, the condensate depletion is still only about
$50$\%, in our case.

\subsubsection{Second-order coherence}

We now look briefly at the second-order coherence function
defined by~\cite{Naraschewski1999a}
\begin{equation}
g^{(2)}(\br,\br') = {\langle \hat{\Psi}^\dagger(\br)\hat{\Psi}^\dagger(\br')
\hat{\Psi}(\br')\hat{\Psi}(\br)\rangle \over
\langle \hat{\Psi}^\dagger(\br)\hat{\Psi}(\br)\rangle
\langle \hat{\Psi}^\dagger(\br')\hat{\Psi}(\br')\rangle}.
\end{equation}
We will consider this function only in the equal-position limit (i.e. when
$\br = \br'$), in which case it may be written as
\begin{eqnarray}
\lefteqn{g^{(2)}(x,x)  = } \label{eq:g2xx} \\
&&1+ {2N_0|\psi(x)|^2[\rho(x)+\kappa(x)]+|\rho(x)|^2+
|\kappa(x)|^2 \over \left[ N_0|\psi(x)|^2 +\rho(x) \right]^2}, \nonumber
\end{eqnarray}
following a similar argument as for the first-order coherence function.  The
second-order equal-position coherence function is of interest because of the
various limits in which exact results are known.  For a pure BEC with perfect
second-order coherence we have $g^{(2)}(x,x) = 1$, whilst for a thermal cloud
of bosons $g^{(2)}(x,x) = 2$~\cite{Dodd1997a,Naraschewski1999a}.  In the
strong-coupling limit in 1D however, the Bose-Fermi mapping theorem predicts
that $g^{(2)}(x,x) = 0$~\cite{Girardeau2001a} for impenetrable bosons.  We
would therefore expect to find a second-order coherence of close to $1$ in the
weakly-interacting limit, which decreases to $0$ as the interaction strength is
increased, in contrast to the higher dimensional case where the coherence
function would increase to $2$ as the non-condensate became significant
\footnote{This is not entirely correct, in fact, since we have ignored the
strong repulsion between atoms at very small distances, which means that
$g^{(2)}(\br,\br')$ must actually vanish for $\br=\br'$ in all dimensions.  The
actual function $g^{(2)}(\br,\br')$ increases quite rapidly, however, as a
function of $|\br-\br'|$ (on the scale of the scattering length), increasing to
values close to those given in the text.  Equation~(\ref{eq:g2xx}) is therefore
the result for an effective Hamiltonian involving a pseudopotential interaction
term, rather than the bare interparticle potential.  This is discussed in more
detail in~\cite{Naraschewski1999a}.}.

\begin{figure}
\psfrag{xlabel}{$x/R_{\rm TF}$}
\psfrag{ylabel}{$g^{(2)}(x,x)$}
\center{
\epsfig{file = 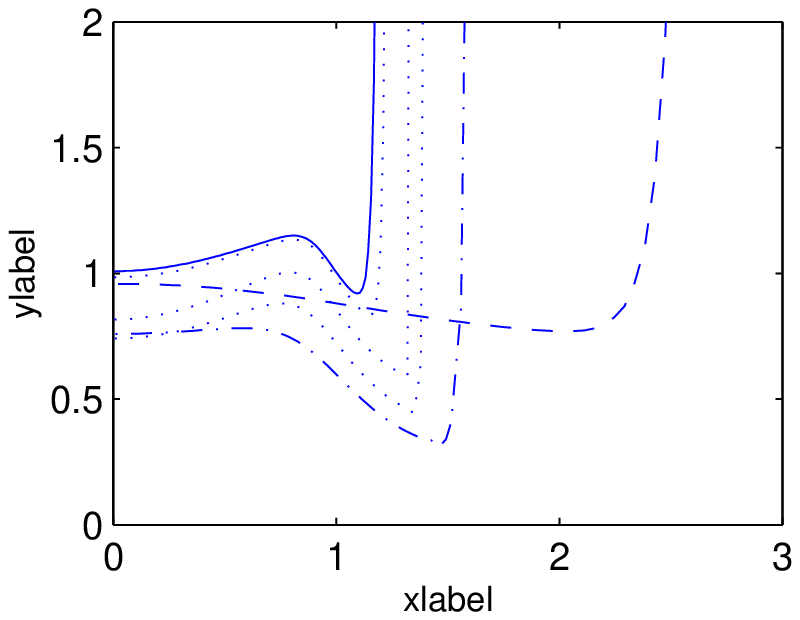,width=8cm}
\Caption{Plot of the normalised equal-position second-order coherence function
$g^{(2)}(x,x)$ obtained from the results of GHFB2 theory for various values of
$V_0$.  The dashed line shows the weakly interacting limit with $\tilde{V}_0 =
0.2$, the dot-dashed line is for $\tilde{V}_0 = 2.4$.  The dotted lines then
correspond to values of $\tilde{V}_0 = 6.5,15,100$ from bottom to top, and the
final solid line is in the strongly interacting limit with $\tilde{V}_0 = 200$
(these values correspond to $\gamma_{\rm trap} = 0.02,0.46,1.6,4.1,31$ and $61$
respectively).  Note that the $x$-axis is scaled in units of the Thomas-Fermi
radius of the system. \label{fig:chap6_GHFB2secondorder}}}
\end{figure}

Figure~\ref{fig:chap6_GHFB2secondorder} shows $g^{(2)}(x,x)$ as calculated from
the results of the GHFB2 theory.  The results can be seen to be close to unity,
as expected, for the weakly-interacting case (and rising to $2$ outside the
region where the condensate exists).  As the interactions are increased,
$g^{(2)}(x,x)$ initially decreases closer to the $g^{(2)}(x,x)=0$ limit
expected.  However, in the very strongly-interacting limit the coherence
function rises again to become greater than $1$.  Thus the GHFB2 theory, whilst
in qualitative agreement with the behaviour of the chemical potential and the
density, does not give the expected results for the coherence of the system in
the impenetrable limit.

\section{Reinterpreting the results}

\begin{figure}
\psfrag{xlabel}{$x$ (units of $\ell_{\rm trap}$)}
\psfrag{ylabel}[Bc]{\hspace{2cm}$N_0|\psi(x)|^2$ (units of $\ell_{\rm trap}^{-1}$)}
\center{
\epsfig{file=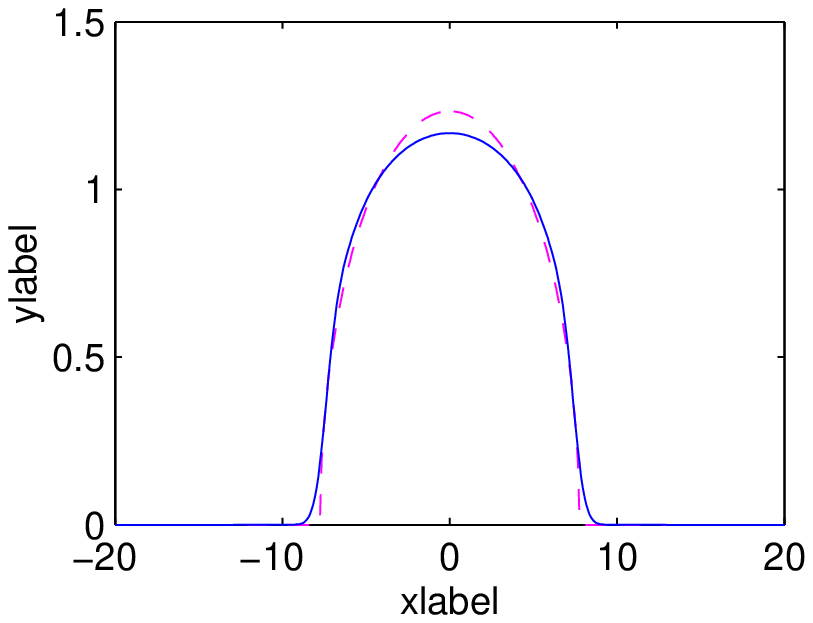,width=8cm}
\Caption{Density due to the condensate population only (solid) resulting from
the GHFB2 theory at $\tilde{V}_0 = 100$ ($\gamma_{\rm trap} = 30$).  The dashed
line again shows the Thomas-Fermi prediction for the exact density for $15$
atoms in the Tonks-Girardeau limit. \label{fig:chap6_GHFB2psidensity}}}
\end{figure}

As seen in the previous section, although the GHFB2 theory seems to go further
than other standard theories in describing the 1D Bose gas into the
strongly-interacting limit, there are significant differences between the
results in the impenetrable limit and the exact results of the Tonks-Girardeau
gas.  However, we will see in this section that the agreement can be improved
significantly by reinterpreting the results in a different manner. The
disagreement with the expected values for the asymptotic value of the chemical
potential and with the shape of the density are both related to the
non-condensate population in the form of $\rho(x)$.  Looking again at
Fig.~\ref{fig:chap6_muvsV} one can see that the chemical potential in the
GHFB2 case rises to an asymptotic value of approximately $15\hbar\omega$. 
Although this result was obtained for a condensate population of $N_0=15$, it
is in excellent agreement with the expected result for a system with $15$ atoms
in total.  By itself this may appear to be a co-incidence; however the shape of
the density resulting merely from the condensate ``wave function'' $\psi(x)$ is
plotted against the Thomas-Fermi exact prediction in
Fig.~\ref{fig:chap6_GHFB2psidensity}, and the two shapes can be seen to agree
quite well. It seems that the best agreement would be midway between that in
Fig.~\ref{fig:chap6_GHFB2psidensity}, and that in
Fig.~\ref{fig:chap6_densities}b (where the total density has been plotted).

This seems to imply that the GHFB2 theory predicts the correct results in the
strongly-interacting limit if one partly ignores the contributions from the
non-condensate (these contributions are sizeable - the non-condensate
population for the results plotted in Fig.~\ref{fig:chap6_GHFB2psidensity} is
about $N_0/2$).  This would have to be combined with a re-interpretation of the
condensate density as the quasi-condensate density and, ultimately, as the
total density.  Some procedure of this kind is clearly required in 1D because
the various theories discussed in the previous section are based on the
existence of a condensate whereas in fact the condensate disappears as $V_0$
increases.

\begin{figure}
\psfrag{xlabel}{$x/R_{\rm TF}$}
\psfrag{ylabel}{$g^{(2)}(x,x)$}
\center{
\epsfig{file = 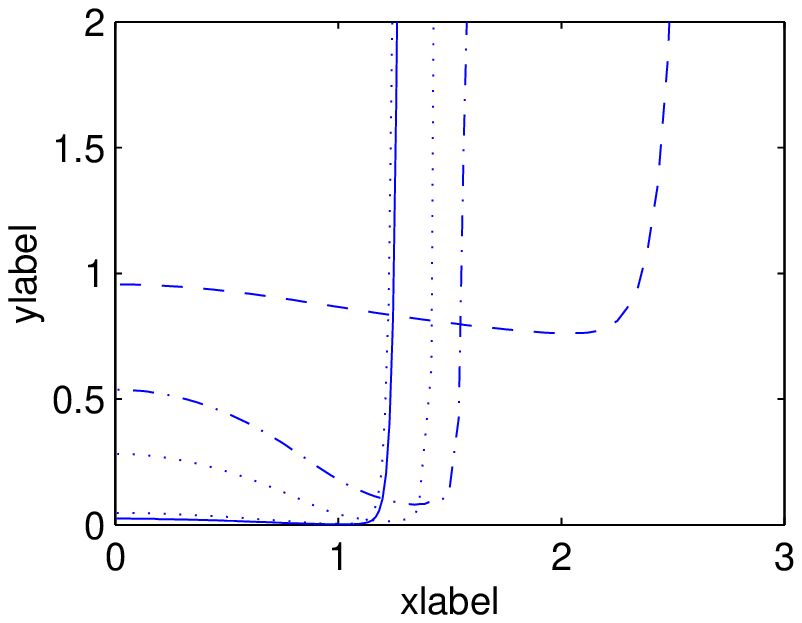,width=8cm}
\Caption{Plot of the normalised equal-position second-order coherence function
$g^{(2)}(x,x)$ obtained from the results of GHFB1 theory with $\rho(x)$
artificially set to zero everywhere, for various values of $V_0$.  The dashed
line shows the weakly interacting limit with $\tilde{V}_0 = 0.2$, the
dot-dashed line is for $\tilde{V}_0 = 3$.  The dotted lines then correspond to
values of $\tilde{V}_0 = 7.5$ and $32$, while the final solid line is in the
strongly interacting limit with $\tilde{V}_0 = 48$ (these values correspond to
$\gamma_{\rm trap} = 0.02,0.65,2.1,11$ and $17$ respectively).  Note that the
$x$-axis is scaled in units of the Thomas-Fermi radius of the system.
\label{fig:chap6_GHFB1zerorhosecondorder}}}
\end{figure}

As another illustration of the good agreement with exact results which can be
obtained with such a reinterpretation, in
Fig.~\ref{fig:chap6_GHFB1zerorhosecondorder} the second-order equal-position
coherence function is plotted with results obtained from the GHFB1 theory with
$\rho(x)$ artificially set to zero (as mentioned earlier, this approach gave
qualitative agreement with the expected behaviour of the chemical potential). 
The results in this case clearly show the behaviour predicted in the previous
section, with $g^{(2)}(x,x)$ decreasing from unity to zero as the interaction
strength is increased.  Such a theory (with $\rho(x)=0$) would result from a
reinterpretation of $N_0|\psi(x)|^2$ as the total density of the system, rather
than merely that of the condensate.   The agreement with the expected behaviour
for $g^{(2)}(x,x)$ can easily be seen from the following argument.  In the
high-$V_0$ limit we have shown that $T_{\rm MB}$ becomes independent of $V_0$. 
From Eq.~(\ref{eq:TMBc6}), this must mean that $\kappa(x) \rightarrow
-N_0\psi_0^2(x)$ in the same limit.  Equation~(\ref{eq:g2xx}) then clearly
shows that the second order coherence function vanishes in the high-$V_0$ limit
if $\rho=0$.

That such a reinterpretation of the condensate is necessary and will lead to
the correct mean-field theory at large $V_0$ can be seen by comparison with the
GPE results in section~\ref{sec:1DGPEdelta}.  If we set $\rho=0$ and
reinterpret $n_0(x)$ as $n_{\rm tot}(x)$ then the GHFB theory reduces to the
earlier GPE theory, the only difference being that $g_{\rm 1D}(x)$ is
calculated from Eq.~(\ref{eq:TMBc6}) instead of
Eq.~(\ref{eq:1Ddeltag}).  Assuming the validity of the local density
approximation, however, these two definitions will lead to the same results. 
We note that a theory has recently been proposed by Anderson \emph{et al.\
}\cite{Andersen2002a,AlKhawaja2002a} which in effect leads to a similar
reinterpretation of the quantities $\rho(x)$, $\kappa(x)$ and $n_0(x)$ by
considering the effects of phase fluctuations to all orders, and this approach
may provide a microscopic justification for this procedure.

\section{Conclusions}

In this paper we have given a form for an approximate many-body T-matrix to
describe the scattering which occurs in a zero temperature 1D Bose gas.  This
approximate T-matrix enables the 1D Bose system to be described by a
Gross-Pitaevskii equation which is easily solved.  We have shown that the
solutions to this equation for a trapped system are in excellent agreement with
the density of a 1D Lieb-Liniger model gas in the two limits in which exact
results are known, and that the density distribution of such a gas changes
markedly as the interactions become stronger.

We have also attempted to push various higher-order theories of BECs from the
weakly-interacting regime into the strongly-interacting regime for the
Lieb-Liniger gas.  We have shown that the theories which neglect many-body
effects on the interactions (BdG and HFB-Popov) fail to provide adequate
results at even quite low interaction strengths.  The GHFB1 theory includes the
many-body T-matrix for condensate interactions, but it also fails to describe
the strongly-interacting limit since it does not account for the
condensate-non-condensate interactions properly and hence the terms involving
$\rho$ become too large.  

However, the GHFB2 theory, which includes many-body effects in the
condensate-non-condensate interactions, gave results which were qualitatively
correct even in the strongly-interacting limit, predicting the correct
behaviour of the chemical potential and density profile.  Quantitative
agreement was not found, but the results show that GHFB2 describes the
strongly-interacting limit to a much better degree than the other mean-field
theories tested.

GHFB2 theory still fails to adequately describe some aspects of the system
however.  The first-order coherence function shows qualitatively the behaviour
expected, but the second-order coherence, while initially showing the  correct
behaviour in the low- to mid-$V_0$ range, deviates from the expected limit at
high-$V_0$.  Furthermore, the quasiparticle excitation spectrum does not go over
to the correct behaviour at high-$V_0$.  

The quasiparticle spectrum can be corrected if the theory is modified to allow
for fluctuations in the many-body T-matrix, as well as in the condensate
density.  This requires a dynamical treatment of the non-condensate.  At the
same time, it seems that the key to pushing the GHFB2 theory into the
high-$V_0$ limit (obtaining the correct coherences and  quantitative agreement
in general) lies in a reinterpretation of the non-condensate density as the
total density, and we have presented some heuristic evidence that this will lead
to the expected results.  

\acknowledgments   
This research was supported by the Engineering and Physical Sciences Research
Council of the United Kingdom, and by the European Union via the ``Cold Quantum
Gases'' network.  S.~M.~thanks Trinity College, Oxford, and the Royal Society
for financial support.  K.~B.~ thanks the Royal Society and the Wolfson
Foundation for support.

\appendix
\section{The many-body T-matrix in 3D}

As an illustration of our approximation of the many-body T-matrix in terms of
the off-shell two-body T-matrix, we show in this appendix that in the
three-dimensional case it reproduces the known results for the many-body
T-matrix.

The 3D case differs from 1D and 2D because the many-body T-matrix is only a
small correction to the two-body T-matrix. As mentioned earlier, the 3D
two-body T-matrix is a constant $U_0 = 4\pi\hbar^2a/m$  to first order in the
low-energy limit.  The first approximation to the 3D coupling parameter is
therefore $g_{\rm 3D} = U_0$.  We can then ask if the method used in this paper
for 1D (and used in reference~\cite{Lee2002a} for 2D) accurately predicts the
effects of the medium on the scattering. 

In~\cite{Lee2002a} we showed that our approximation for the many-body T-matrix
in 3D is
\begin{equation}
g_{\rm 3D} = \bra{0}T_{\rm MB}(E=0)\ket{0} 
\approx \bra{0}T_{\rm 2B}(E=-{16\over \pi^2}\mu)\ket{0},
\label{eq:g3D1}
\end{equation}
and the off-shell two-body T-matrix for a gas of hard-spheres of radius $a$ was shown in~\cite{Morgan2002a} to be
\begin{eqnarray}
\lefteqn{\bra{{\bf 0}} T_{\rm 2B}^{\rm 3D}(E) \ket{{\bf 0}} = } \nonumber \\
&&\left\{ \begin{array}{ll}
U_0 
\left[ 1-i\sqrt{mE \over \hbar^2}a-\frac{1}{3}{mE\over \hbar^2}a^2 \right], 
& \mbox{for } E>0, \\
U_0 \left[ 1+\sqrt{m|E| \over \hbar^2}a-
\frac{1}{3}{mE\over \hbar^2}a^2 \right], & \mbox{for } E<0,
\end{array} \right.
\end{eqnarray}

Combining these equations gives (to first order in the small parameter $k_\mu a
= \sqrt{m\mu a^2/\hbar^2}$) 
\begin{equation}
g_{\rm 3D} = U_0 \left[ 1 + \left( {16 \mu m a^2 \over \pi^2\hbar^2}
\right)^{1/2} \right] + \mbox{O}[(k_{\mu}a)^2] .
\end{equation}
We can rewrite this in terms of the density $n_0$, since in a homogeneous
system $\mu = g_{\rm 3D}n_0$ to leading order.  Making this substitution and
solving for $g_{\rm 3D}$ gives the result
\begin{equation}
g_{\rm 3D} = U_0 \left[ 1 + {8 \over \sqrt{\pi}}(n_0a^3)^{1/2} \right] +
\mbox{O} (n_0a^3),
\end{equation}
where we have expanded in terms of the dilute gas parameter $n_0a^3$, which is
small for a dilute gas BEC.  This result agrees with other
solutions~\cite{Shi1998a,Morgan1999a} for the many-body T-matrix in 3D
homogeneous systems.  At $T=0$ we can, therefore, account for many-body effects
on particle scattering using an off-shell two-body T-matrix.

\bibliography{1Dpaper_bibliography}

\end{document}